\documentclass[final,3p,sort&compress,times]{elsarticle}
\usepackage{graphicx}
\usepackage{dcolumn}
\usepackage{bm}
\usepackage{upgreek}
\usepackage{hyperref}
\usepackage{amsmath}
\usepackage{multirow}
\usepackage{mdframed} 
\usepackage{multicol} 
\usepackage{nomencl} 
\makenomenclature
\usepackage{hyperref}

\journal{International Journal of Heat and Mass Transfer}

\begin{document}

\begin{frontmatter}

\title{Effect of startup modes on cold start performance of PEM fuel cells with different cathode flow fields}

\author[TJU]{Wenzhe Zhang}
\author[TJU]{Xingxiao Tao}
\author[TJU]{Qifeng Li}
\author[TJU]{Kai Sun}
\author[TJU,LU]{Rui Chen}
\author[TJU,NIEPES]{Zhizhao Che\corref{cor1}}
\ead{chezhizhao@tju.edu.cn}
\author[TJU,NIEPES]{Tianyou Wang\corref{cor1}}
\ead{wangtianyou@tju.edu.cn}
\cortext[cor1]{Corresponding authors.}

\address[TJU]{State Key Laboratory of Engines, Tianjin University, Tianjin, 300350, China.}
\address[NIEPES]{National Industry-Education Platform of Energy Storage, Tianjin University, Tianjin, 300350, China}
\address[LU]{Department of Aeronautical and Automotive Engineering, Loughborough University, Loughborough LE11 3TU, United Kingdom}

\begin{abstract}
Proton Exchange Membrane Fuel Cell (PEMFC) is widely recognized for its cleanliness and high efficiency, but is still facing challenges in cold environments. At low temperatures, the formation of ice and repeated freezing/thawing cycles may cause cell performance reduction and irreversible degradation. The cathode flow field of PEMFCs has a significant effect on the performance. In contrast to the conventional ``channel-ridge'' flow field, the metal foam has the advantages of excellent pre-distribution of gases and water drainage, which make it a promising candidate for the cold start. This paper examines the cold start of PEMFCs with metal foam flow field (MFFF) and serpentine flow field (SFF), and the influence of constant current mode, constant voltage mode, and ramping current mode is investigated experimentally through performance test and electrochemical characterization. The results show that lowering the voltage and increasing the current can enhance the cold-start performance of fuel cells. The MFFF fuel cell has superior cold start performance compared to the SFF fuel cell under the constant voltage mode of 0.3 V. Furthermore, the variable current mode is developed by considering the distinct properties of heat and water production during various phases, and the results indicate that increasing the current density at the unsaturated stage leads to an elevated rate of heat production and a reduced rate of water production, which can improve the cold start of PEMFCs.
\end{abstract}

\begin{keyword}
\texttt {
Cold start \sep
Metal Foam Flow Field \sep
Water freezing \sep
Startup mode \sep
Proton exchange membrane fuel cell
}
\end{keyword}

\end{frontmatter}

\section{Introduction}\label{sec:1}
Proton Exchange Membrane Fuel Cell (PEMFC) is a promising technology that demonstrates exceptional efficiency and environmental friendliness in the realm of energy conversion \cite{pramuanjaroenkij23, yang21}, and has great application prospects \cite{lin17, wang17}. However, the operation of PEMFCs in low-temperature environments is still facing great challenges \cite{lin21, luo18}. In instances where the temperature falls below the point of freezing, water produced by the electrochemical reaction within the PEMFCs may freeze, which will have an adverse influence on the operational efficiency \cite{hou11, yang21assistedheating}. Hence, delaying the icing of fuel cells during startup and preventing irreversible structural damage due to icing is crucial to enhancing the performance of PEMFCs \cite{tabe12, Zhang2024}.

Many investigations have been conducted regarding the startup of PEMFCs in cold temperature conditions, including the water transport \cite{jiao11}, heat transfer \cite{chippar12}, startup strategy \cite{hu21stack}, etc. The water transport and phase change play crucial roles in influencing the cold start performance of PEMFCs, and the temperature variation affects the conversion of water and ice. Regarding water production and its phase change, Ge et al. \cite{ge07} designed a transparent PEMFC to facilitate the observation of the water state within the catalyst layer (CL). They found that water is present in both ice and vapor phases within the cathode CL at 0.02 A/cm$^2$ when the startup temperature is -5 $^\circ$C; when the startup temperature is raised to -3 $^\circ$C, water in the liquid state becomes visible on the CL. Misher et al. \cite{mishler12} investigated the location of ice formation in PEMFCs using neutron imaging, and the results revealed that the maximum ice thickness was observed on the cathode side of the membrane electrode assembly (MEA). Jiang et al. \cite{jiang20} formulated a one-dimensional (1D) model to simulate the cold start of PEMFCs, and the results indicated an uneven distribution of water and ice within the CL. Yao et al. \cite{yao18} established models of freezing/thawing in gas diffusion layer (GDL) and CL by considering non-equilibrium water transfer in MEA, and found that the ability of cold start is influenced by the water storage capacity of the electrolyte in the cathode CL. When the cold start fails, the primary accumulation of ice is predominantly observed in the cathode CL. By simulations of real-scale PEMFC geometries, Jo et al. \cite{jo15} show that the ice accumulation regions are different for different stoichiometric ratios. At low stoichiometry, ice accumulates near the edge of the cell, while at high stoichiometry, ice preferentially accumulates at the cathode inlet region. A mathematical model was developed by Min et al. \cite{min22} to calculate the distribution of temperature and the heat exchange during the cold start. Sundaresan et al. \cite{sundaresan05} developed a layered model of PEMFC stacks to investigate the effect of endplates on cold starts, and found that the performance can be enhanced by heating the end plates, using metal-based materials, and reducing the thickness of the end plates. Luo et al. \cite{luo13} established a three-dimensional (3D) model for PEMFC stacks. The results show that an increase in the number of cells leads to the voltage decays at a slower rate and the temperature increases at a faster rate due to the decreased rate of ice generation. Ultimately, the temperature of the PEMFC stack exhibits a higher level when the startup fails. Ishikawa et al. \cite{ishikawa08} established a system that uses visible and infrared lights to image the internal interface of fuel cells and found that at the moment of performance degradation, the GDL-MEA interface experiences the diffusion of solidification heat.

The cathode flow field determines the reactant gas distribution and water drainage \cite{santamaria13,Gao2023}, so it plays a crucial role in influencing the performance of PEMFCs \cite{zhu21,ValentinReyesi2022}. Liao et al. \cite{liao21} conducted a comparative analysis of the low-temperature startup of PEMFCs with a zigzagged channel and a straight channel flow field, and found that the zigzagged channel facilitated the transport of reactants in the channel region, which leads to a more homogeneous distribution of electric current, concentration of oxygen, and formation of ice. The influence of channel geometry on the cold start of PEMFCs was studied by Hu et al. \cite{hu21}. They found that the change of the flow channel geometry had significant effects on water content and current density, but minor effects on temperature and ice volume fraction. It is essential to know that the traditional flow field is designed with an alternating ``channel-ridge'' structure \cite{kim18, suo22}. The ``channel-ridge'' structure could result in non-uniform dispersion of reactant gas, and the generated water will accumulate under the ridge, causing the porous electrode to freeze in sub-zero environments. Metal foam flow fields (MFFF) have been proposed for fuel cells due to their porous structure, high mass and thermal conductivities, and good water drainage capacity \cite{awin19, carton17, tseng12}. Jo et al. \cite{jo18} found that MFFF exhibits a more homogeneous distribution of membrane water and current density in comparison to the parallel serpentine flow field. Afshari et al. \cite{afshari17} found that compared with parallel flow field and partial baffle parallel flow field, the MFFF can enhance the oxygen concentration and current density at the cathode CL, and improve the uniformity of their distribution. Shin et al.\cite{shin18} found that the maximum power density of fuel cells can be improved by using appropriate pore sizes of metal foam in comparison to fuel cells with a serpentine flow field (SFF). Several investigations have been conducted on the cold start of PEMFCs with MFFF. Huo et al. \cite{huo17} performed experiments on PEMFCs with MFFF and parallel flow field. The results indicated that the MFFF fuel cell has enhanced cold start performance due to its superior mass and heat transfer uniformity. Experimental studies by Xie et al. \cite{xie19} showed that fuel cells with MFFF exhibited higher output voltage and better ice storage capacity during cold start.

The startup mode can significantly influence the cold start performance of PEMFCs \cite{amamou16}. Common startup modes include constant current, constant voltage, and constant power modes. Tao et al. \cite{tao21} established a 1D transient model and investigated the water and heat transfer inside fuel cells under different startup current densities. The results showed that an elevated startup current leads to a more rapid increase in the fuel cell temperature, but at the same time, the ohmic polarization also increases, resulting in a failure to cold start. Zang et al. \cite{zang20} introduced a 1D transient model in their study of the influence of constant current mode. The findings indicate that the success of the startup depends on a specific threshold of current density. Lin et al. \cite{lin14} examined the cold start performance in the constant voltage mode, and found that a decrease in the startup voltage leads to an increased generation of heat, which in turn facilitates the heating of the PEMFC. Jiang et al. \cite{jiang08} explored the influence of constant voltage mode on cold start, which exhibits a significant enhancement in current density, resulting in more heat generation and faster temperature rise of PEMFCs. They found that in comparison to the constant current mode, the constant voltage mode offers more benefits when the membrane is dried with sufficient gas purge. The investigation conducted by Yang et al. \cite{yang21ConstantVoltage} investigated the influence of startup voltage experimentally. The results showed that a lower initial voltage will increase the heat production rate. Luo et al. \cite{luo14} developed a constant power mode, and found that the fuel cell under the constant power mode may stop operating prior to the total coverage of the cathode CL by ice because the fuel cell cannot provide the corresponding output power. Jiang et al. \cite{Jiang10} explored the influence of ramping current modes with different initial current densities and current slopes, and found that larger initial currents and higher current slopes would accelerate fuel cell heating. Lei et al. \cite{lei22} introduced a gradually changing current density mode. Under the proposed mode, the fuel cell exhibits an increased heating rate, a decreased output voltage, and an improved equilibrium between water generation and freezing. Du et al. \cite{du14} established a 1D cold start model and proposed the maximum power mode, i.e., the performance of the stack at each moment was calculated by simulation, and the applied load is the maximum power at that moment. The study revealed that the maximum power mode offers a better equilibrium between the generation of heat and the formation of ice.

Although numerous investigations have been conducted on the cold start of PEMFCs with traditional ``channel-ridge'' flow fields, there is a lack of research related to the cold start of the MFFF fuel cells, especially the influence of startup mode is not clear. This study investigates the cold start performance of PEMFCs through a series of experiments, focusing on the influence of startup mode on the MFFF fuel cells. This study comparatively examines the cold start characteristics of fuel cells with metal foam flow field (MFFF) and serpentine flow field (SFF). We focus on the influence of different cold start modes on the cold start performance of fuel cells. We analyze many characteristics of fuel cell cold start experimentally, including the current and voltage, temperature evolution, heat production, water production, operation time, final temperature rise, and high-frequency resistance (HFR). The influence of constant current, constant voltage, ramping current, and variable current startup modes are compared.

\section{Experimental}\label{sec:2}
\subsection{Experimental setup}\label{sec:2.1}
\begin{figure}
  \centering
  \includegraphics{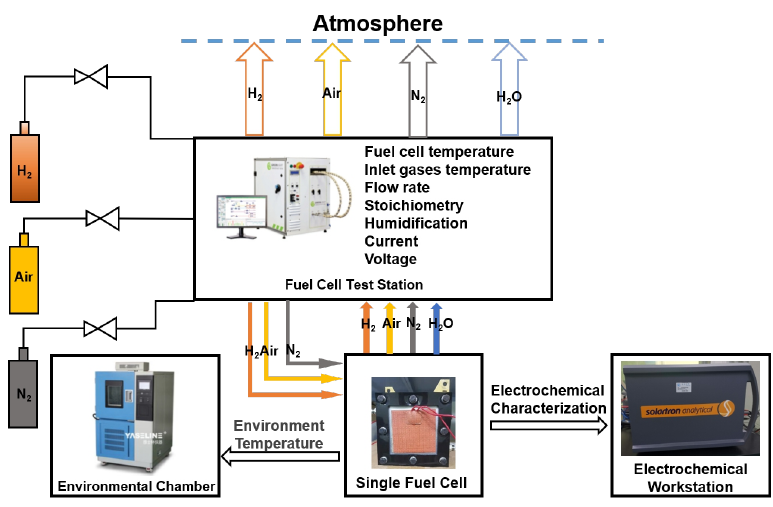}
  \caption{Schematic diagram of fuel cell cold start experiment system.}\label{fig:01}
\end{figure}

The experimental configuration is depicted in Figure \ref{fig:01}. A test bench was used to set the working parameters and collect data on fuel cells, including anode/cathode inlet flow rates, inlet humidity, fuel cell temperature, output voltage, working current, etc. The ambient temperature was controlled with an environmental chamber, and an electrochemical workstation was used to characterize the HFR. T-type thermocouples were used to measure the temperature of the PEMFC by inserting them into the bipolar plates (BP).

\begin{figure}
  \centering
  \includegraphics[scale=0.8]{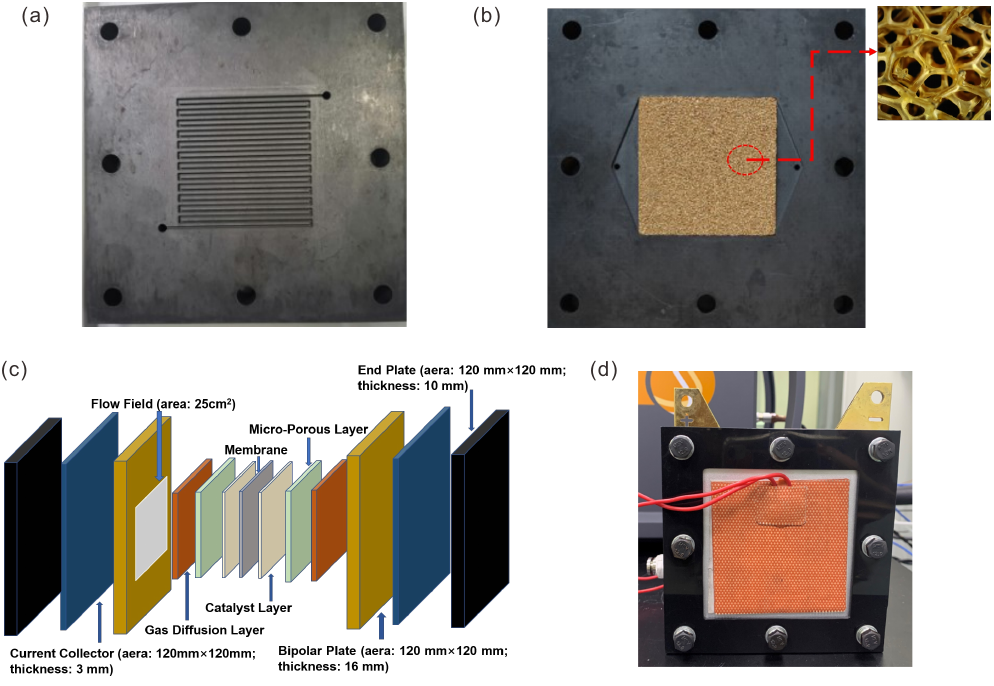}
  \caption{(a) SFF. (b) MFF. (c) Diagram of the fuel cell structure. (d) Assembled fuel cell.}\label{fig:02}
\end{figure}

In this study, two PEMFCs with different cathode flow fields were designed (see Figure \ref{fig:02}). The MFFF fuel cell uses a gold-plated nickel metal foam embedded in a graphite flow field plate, while the SFF fuel cell uses a serpentine cathode flow field made directly on a graphite plate. The nickel metal foam initially had 75 pore per inch (PPI), a porosity of 98\% and a thickness of 3 mm, and it was compressed to a thickness of 1 mm to be used as the cathode flow field. The area of the metal foam corresponds to 25 cm$^2$, which aligns with the area of the membrane electrode, and is consistent with that in the literature (see Refs.\ \cite{tseng12, shin18, WOS:000637968700002, WOS:000540378500076, WOS:000773396900003}). Meanwhile, the same anode serpentine flow field was adopted for both fuel cells. The cross-sectional dimensions of both the channels and ridges in the SFF are 1 mm $\times$ 1 mm. The total length of the channel is 125 cm, and the effective reaction area is 25 cm$^2$. Except for the cathode flow field, all parameters of the two fuel cells are the same. Commercial perfluorinated sulfonic seven-layer MEAs (WUT-HyPower) were used in the experiment, and their anode and cathode have platinum loadings of 0.1 and 0.4 mg/cm$^2$, respectively. Throughout the experiment, the ambient temperature in the room was about 25 $^\circ$C. We increased the length of the intake pipe of the fuel cell in the ambient chamber as much as possible, which allowed the gas temperature to decrease before entering the fuel cell.

\subsection{Fuel cell characterization by High Frequency Resistance}\label{sec:2.2}
To monitor the working condition of the fuel cell, the High Frequency Resistance (HFR) of the fuel cell was monitored by the electrochemical workstation. The HFR method is an Electrochemical Impedance Spectroscopy (EIS) test method specifically employed to assess the real-time ohmic resistance of fuel cells \cite{tabe12, WOS:000245093800034}. The ohmic impedance of fuel cells consists of several components, including membrane resistance, catalyst layer electronic resistance, catalyst layer ionic resistance, gas diffusion layer ohmic resistance, flow field plate ohmic resistance, collector plate ohmic resistance, and interface contact resistance \cite{yang21}. The electrical resistance of the catalytic layer is significantly smaller in comparison to the ionic resistance and can be practically disregarded. The membrane resistance and ionic resistance of the catalyst layer vary based on the water content and temperature of the cell, whereas the contact resistance and other resistances remain unaffected by the water content. Consequently, HFR is employed to characterize the variation in water content in fuel cells.
\subsection{Cold Start Experiment Procedure}\label{sec:2.3}
To improve the catalyst activity and optimize the performance of the PEMFCs, a pre-activation procedure was performed on the PEMFCs prior to the cold start experiment \cite{WOS:000295602400001, WOS:000470045800048}. In this process, hydrogen at a flow rate of 1 SLPM (Standard Liter Per Minute) was introduced into the anode, while air at 3 SLPM was introduced into the cathode. The humidity of the gases was set at 100\%. The PEMFCs and inlet gases were both kept at a constant temperature of 70 $^\circ$C. Generally, the fuel cells were continuously operated for 6 hours near the peak power density, so that the PEMFCs could achieve a stable output performance.
The cold start experiment includes three stages: pre-condition, purge and freezing, and cold start test.
\begin{enumerate}
  \item Pre-condition: Firstly, we operated the fuel cell under normal working conditions for a while (i.e., 30 min at a constant voltage of 0.6 V) \cite{WOS:000403381500046, WOS:000416498700012}. The purpose is to confirm that the last cold start test did not result in significant degradation to the PEMFC and the output performance is still high enough. The pre-condition is also to simulate the internal water distribution after a period of operation.
  \item Purging and freezing: After the pre-condition, dry nitrogen gas was introduced into the anode and cathode at a flow rate of 2 SLPM. The purpose is to eliminate the water generated during the pre-condition. The initial water content of the membrane is a critical variable that influences the performance of PEMFCs, so the purging process should be strictly controlled. We adopted the method reported in the literature \cite{yang21} to monitor the change in membrane water content during the purging stage by measuring the HFR. The HFR of fuel cells mainly consists of $R_s$, which is independent of the water content (including the electrical resistance, CL resistance, and contact resistance), and $R_m$, which is related to the membrane water content. The value of $R_s$ of the fuel cell can be determined by fitting HFR via equilibrium purge measurements, which establish a thermodynamic equilibrium between the hydration of the membranes and the relative humidity of the purge gases. Then the expected HFR for purging can be calculated based on the preset membrane water content. Next, in the purging stage, we used intermittent purging until the HFR reached our calculated value. In intermittent purging, the introduction of dry nitrogen into the fuel cell results in a significant rise in the HFR. Upon ceasing the nitrogen supply, the water within the cell will be redistributed to rehydrate the membrane, resulting in a decrease in the HFR. However, the post-decrease HFR value will be larger than the initial value before the purge. Then we repeated the aforementioned steps until the HFR of the fuel cell achieved the predetermined value. After the purging process, the PEMFC was put into the environmental chamber with the initial temperature of cold start until the PEMFC reached the initial temperature. The freezing process took about 3 hours.
  \item Cold start test: After the PEMFC reached the preset initial startup temperature, the anode of the PEMFC was supplied with 1 SLPM dry hydrogen, while the cathode was supplied with 3 SLPM dry air. The temperatures of the dry hydrogen and air were set the same as the initial startup temperature. The data collection procedure was conducted for 1200 s or until the cold start procedure fails (i.e., the output voltage falling below 0.01 V or the working current density below 0.02 A/cm$^2$).
\end{enumerate}
After the cold start test, the next round of the experiment was carried out from the pre-condition. During the cold start process, we define the instant when the electronic load is applied as $t = 0$. The HFR measurement was performed by using an electrochemical workstation. The electrochemical workstation continuously records the high-frequency resistance (HFR) to monitor the hydration and icing in the fuel cells. The disturbance current is set to 100 mA and the frequency is 1000 Hz. Samples were taken at intervals of 2 s (purging), 1 min (cooling), and 1 s (cold start), respectively.

\subsection{Data analysis}\label{sec:2.4}
A successful cold start of PEMFC is primarily affected by two key factors, namely water production rate and heat production rate. To attain a successful cold start, it is necessary to maintain a high heat production rate but a low water production rate. The total heat production ($Q_\text{Heat}$, J/cm$^2$) and water production ($m_\text{H2O}$, mg/cm$^2$) can be calculated by the following equation, respectively.
\begin{equation}\label{eq:01}
  {{Q}_{\text{Heat}}}=\int{({{E}_{h}}-{{V}_\text{cell}})}Idt\approx ({{E}_{h}}-{{V}_\text{cell}})\int{Idt}
\end{equation}
\begin{equation}\label{eq:02}
  {{m}_{{{\text{H}}_{\text{2}}}\text{O}}}\text{=}\frac{{{M}_{{{\text{H}}_{\text{2}}}\text{O}}}}{2F}\int{Idt}
\end{equation}
where $E_h$ represents the ideal thermal potential (1.48 V), $M_\text{H2O}$ represents the molecular weight of water, $F$ is the Faraday constant (96478 C/mol), $V_\text{cell}$ represents the output voltage, and $I$ represents the instantaneous current density.

In practical applications of fuel cell systems, it is common practice to store hydrogen in a high-pressure tank, and there is no need for a gas pump to supply hydrogen to the anode. But on the cathode side, it is generally necessary to use an air compressor to introduce air, therefore the pump power loss should be considered in the net power density. The pump power loss can be calculated as:
\begin{equation}\label{eq:03}
{{P}_{\text{pump}}}=\frac{{{m}_{\text{air}}}{{c}_{p}}T}{A\zeta }\left[ {{\left( \frac{{{p}_{\text{inlet}}}}{{{p}_{\text{atm}}}} \right)}^{\frac{k-1}{k}}}-1 \right]
\end{equation}
where $P_\text{pump}$ is the density of pump power loss (W/cm$^2$); $m_\text{air}$ is the mass flow rate of air (kg/s); $c_p$ is the isobaric specific heat capacity (J/(kg$\cdot$K)); $T$ is the absolute temperature of air (K); $A$ is the effective reaction area (cm$^2$); $p_\text{inlet}$ is the inlet pressure of the cathode (Pa); $p_\text{atm}$ is the atmospheric pressure (Pa); $k$ is the adiabatic index of the air, which generally takes 1.4; $\zeta $ is the isentropic coefficient of the air compressor. The air compressor in this experiment is a twin-screw rotary air compressor, and the isentropic coefficient is taken as 0.7 \cite{Boettner02}. The net power density $P_\text{net}$ is \cite{cooper16}:
\begin{equation}\label{eq:04}
  {{P}_{\text{net}}}={{P}_{\text{raw}}}-{{P}_{\text{pump}}}
\end{equation}
where $P_\text{raw}$ is the total power density (W/cm$^2$).

\section{Results and discussion}\label{sec:3}
\subsection{Performance of PEMFC under room temperature conditions}\label{sec:3.1}
\begin{figure}
  \centering
  \includegraphics{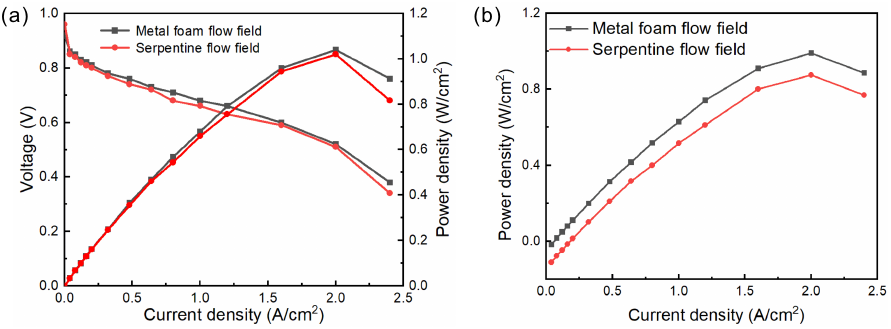}
  \caption{Performance of the MFFF and SFF fuel cells at room temperature: (a) Polarization curves and power density curves; (b) Net power density curve.}\label{fig:03}
\end{figure}

Before conducting the cold start experiments, we characterize the performance of the PEMFC under room temperature conditions. According to the polarization curve in Figure \ref{fig:03}a, the MFFF fuel cell demonstrates a slightly higher output voltage in comparison to the SFF fuel cell across various current densities. The maximum power density of the MFFF fuel cell exhibits a slight increase compared with the SFF fuel cell. The better performance of the MFFF fuel cell can be attributed to the porous structure of the metal foam, which enables air to be more evenly distributed in the PEMFC. The capability of the metal foam to uniform air distribution can promote the electrochemical reaction, which leads to increased output voltage \cite{huo17, xie19}. The serpentine flow field is limited by the ``channel-ridge'' structure. Only the channel part is used for gas transfer, and the oxygen concentration loss is larger. Meanwhile, the porous structure can provide more microchannels to enhance water drainage. In contrast, water in the SFF fuel cell will easily accumulate beneath the ridge and present challenges for effective discharge from the PEMFC, particularly at high current densities. This accumulation of liquid water has the potential to cause water flooding within the PEMFC and reduce fuel cell efficiency.

In the experiment, constant flow rates are set at the inlets. The flow rate of air is 2 SLPM, while the absolute cathode inlet pressures of the MFFF and SFF fuel cells are 124.1 kPa and 175.8 kPa, respectively. Hence, according to Eq. (\ref{eq:03}), the pump power loss densities of the MFFF and SFF fuel cells are 0.0506 and 0.145 W/cm$^2$, respectively. The power density in Figure \ref{eq:03}a represents the overall power density of the MFFF and SFF fuel cells. It is calculated by multiplying the current density and voltage, both of which are directly measured. The net power densities of the PEMFCs are shown in Figure \ref{fig:03}b, representing the net power density of the MFFF and SFF fuel cells. It is obtained by subtracting the pump power density from the overall power density. The net power densities of the PEMFCs are depicted in Figure \ref{fig:03}b. In comparison to the SFF fuel cell, the MFFF fuel cell has a greater net power density, with a maximum net power density increase of approximately 13\%. This is mainly because the SFF only has a single channel and its pressure drop is large, which is a disadvantage in practical applications. Therefore, if considering the net power, it is evident that the MFFF fuel cell exhibits significantly enhanced output performance compared with the SFF fuel cell.

\subsection{Cold start performance of fuel cells under different startup modes}\label{sec:3.2}
In this study, the cold start performance of the MFFF and SFF fuel cells under different startup modes is investigated. Based on the constant voltage, constant current, and ramping current modes, we also designed variable current modes. The effects of starting modes are investigated at different initial startup temperatures.

\subsubsection{Constant current mode}\label{sec:3.2.1}
\begin{figure}
  \centering
  \includegraphics[width=\columnwidth]{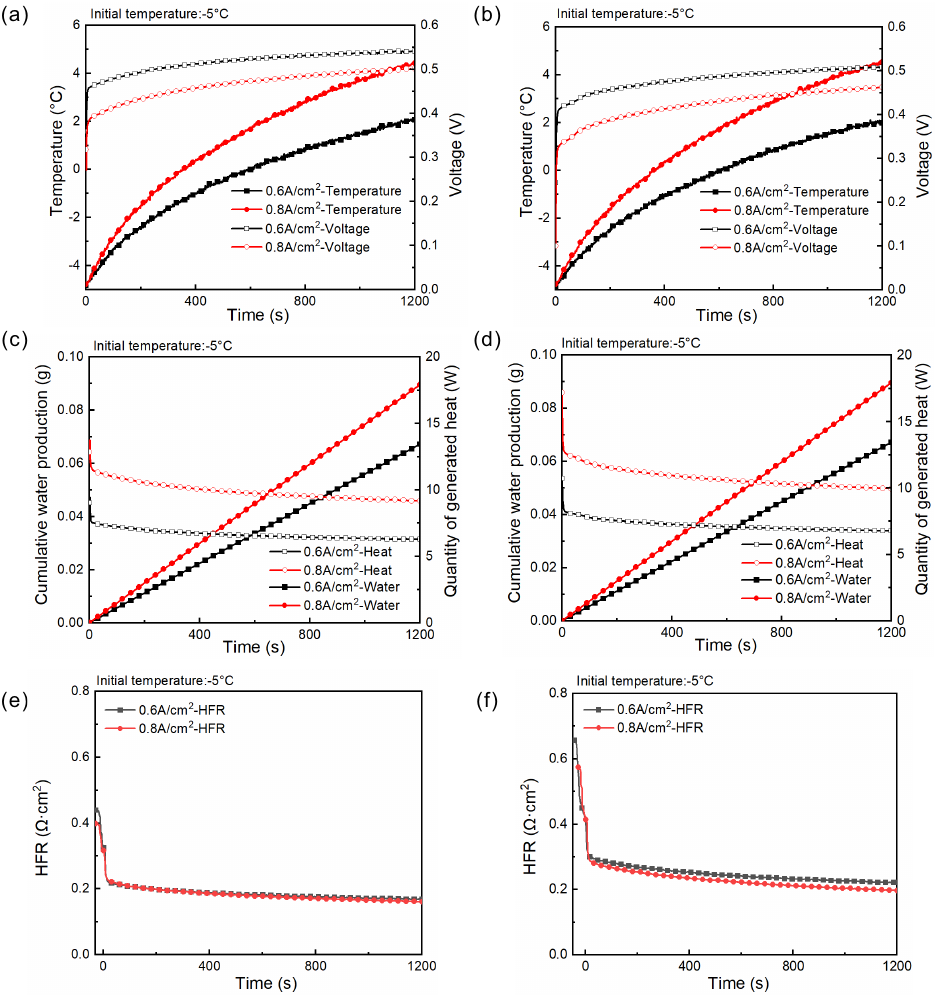}
  \caption{Time variation of parameters of the PEMFC under the constant current mode when the initial startup temperature is -5 $^\circ$C: (a, c, e) MFFF fuel cell; (b, d, f) SFF fuel cell. $t = 0$ in the figure corresponds to the instant when the electronic load is applied.}\label{fig:04}
\end{figure}

During the experiment, we conducted tests at two specific current densities, namely 0.6 and 0.8 A/cm$^2$, under the constant current mode. Lowering the initial current densities will prolong the time for the rise of the fuel cell temperature. Hence, it will take more time for a successful cold start of the fuel cell. Moreover, it is difficult to achieve very high current densities for fuel cells at low temperatures. The MFFF and SFF fuel cells were able to successfully start at the initial temperature of -5 $^\circ$C under constant current modes. During the cold start, the fuel cell undergoes a progressive increase in temperature and output voltage. This phenomenon mostly occurs due to the production of heat by the fuel cells during the reaction, which leads to its self-heating. The elevated temperature results in increased membrane conductivity, expedited electrochemical reaction, and progressively increased output voltage. The increase rates of temperature and voltage steadily decline due to a decrease in the heat production rate, which is primarily caused by a steady decrease in the ohmic resistance. Comparing the two current densities, it is evident that a higher current density results in a more rapid heating rate in the fuel cell. This is mostly due to the fuel cell's increased heat generation rate. Figures \ref{fig:04}e and \ref{fig:04}f depict the variation of the HFR during the cold start phase. The variation in the HFR before $t =0$ (when the load is imposed) is caused by the introduction of dry air and hydrogen into the fuel cell. Upon the application of the load ($t =0$), Figures \ref{fig:04}e and \ref{fig:04}f clearly demonstrate a significant fall in the HFR, followed by a gradual decline. This observation suggests that the porous electrode is not entirely obstructed by ice. Moreover, the final HFRs of the two fuel cells at the current density of 0.8 A/cm$^2$ are lower compared to those at the current density of 0.6 A/cm$^2$. These findings indicate that an increased current density leads to a higher production of water, which is consistent with the observed trends in Figures \ref{fig:04}c and \ref{fig:04}d.

\begin{figure}
  \centering
  \includegraphics[width=\columnwidth]{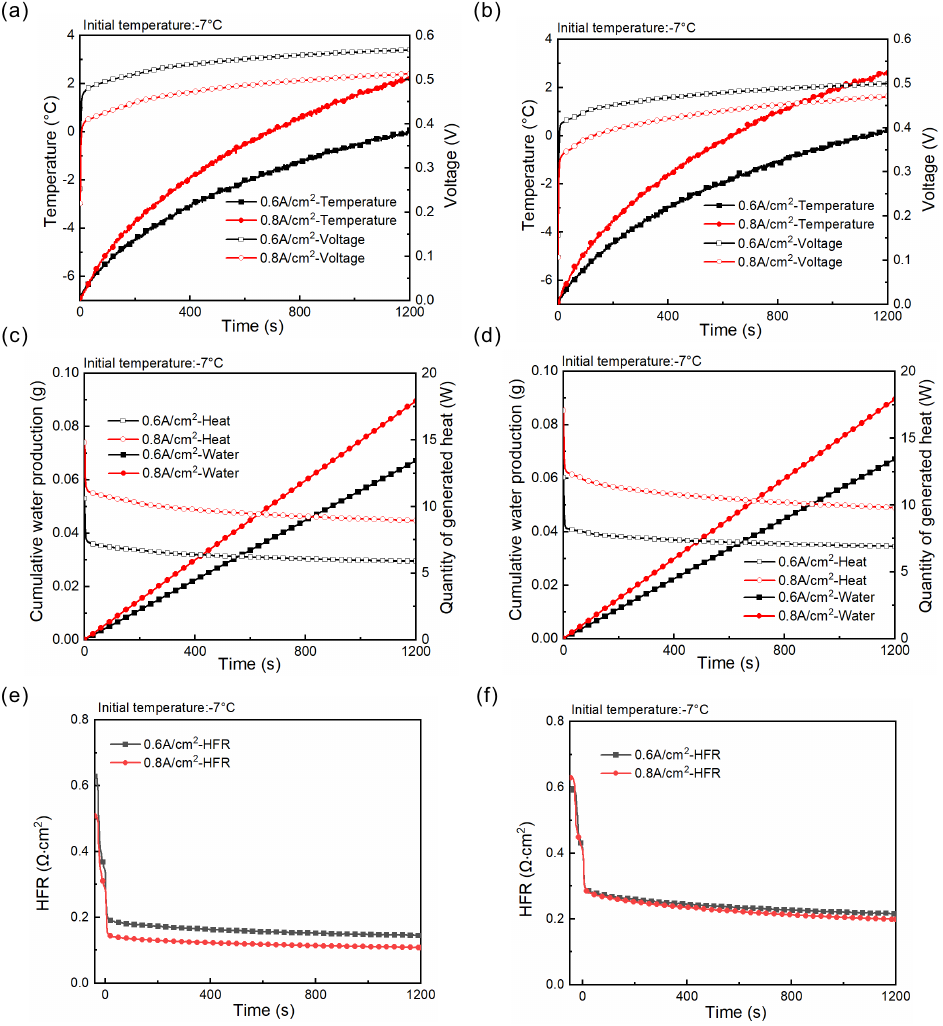}
  \caption{Time variation of parameters of the PEMFC under the constant current mode when the initial startup temperature is -7 $^\circ$C: (a, c, e) MFFF fuel cell; (b, d, f) SFF fuel cell. $t = 0$ in the figure corresponds to the instant when the electronic load is applied.}\label{fig:05}
\end{figure}

At an initial temperature of -7 $^\circ$C, the MFFF and SFF fuel cells were able to start successfully under the constant current mode of 0.6 and 0.8 A/cm$^2$. It took 1225 s and 703 s for the MFFF fuel cell to reach a temperature of 0 $^\circ$C at the current density of 0.6 and 0.8 A/cm$^2$, respectively. Similarly, for the SFF fuel cell, it took 1140 s and 634 s, respectively. Figures \ref{fig:05}e and \ref{fig:05}f demonstrate that, under the constant current mode of 0.6 and 0.8 A/cm$^2$, the HFR did not exhibit a sudden increase during the cold start, suggesting there was no evident icing during the cold start. Both the MFFF and SFF fuel cells exhibit a lower final value of the HFR at 0.8 A/cm$^2$ than at 0.6 A/cm$^2$ in the constant current mode, similar to that at the initial temperature of -5$^\circ$C, indicating that the fuel cell generates a greater amount of water when operating at a high current density.

\begin{figure}
  \centering
  \includegraphics[width=\columnwidth]{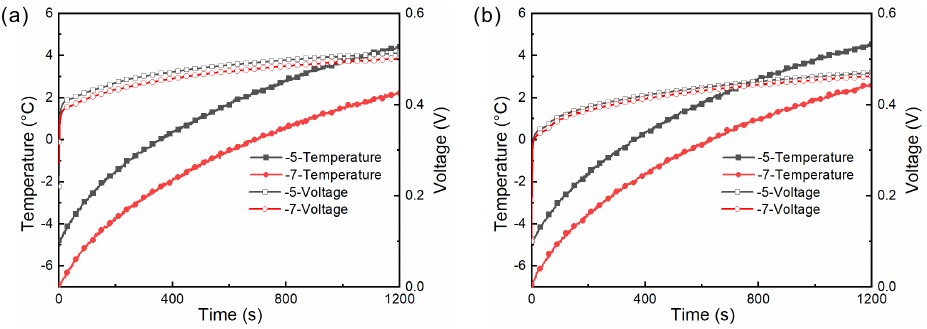}
  \caption{Temperature and voltage variations under constant current mode at 0.8 A/cm$^2$ for different initial startup temperatures: (a) MFFF fuel cell; (b) SFF fuel cell.}\label{fig:06}
\end{figure}
The performance of the fuel cells under constant current mode at -5 $^\circ$C and -7 $^\circ$C is compared in Figure \ref{fig:06}. The time of temperature rise of the MFFF and SFF fuel cells is longer at the initial temperature of -7 $^\circ$C compared with the initial temperature of -5 $^\circ$C, because the lower initial startup temperature needs to consume more heat for fuel cell heating. Comparing the two fuel cells, it is evident that the MFFF fuel cell exhibits a greater operating voltage in comparison to the SFF fuel cell. According to Eq.~(\ref{eq:01}), the heat production rate of the SFF fuel cell surpasses that of the MFFF fuel cell. Meanwhile, the water production rate of the PEMFCs is the same. Therefore, the SFF fuel cell reaches 0 $^\circ$C in a shorter time than the MFFF fuel cell.

\begin{figure}
  \centering
  \includegraphics[width=\columnwidth]{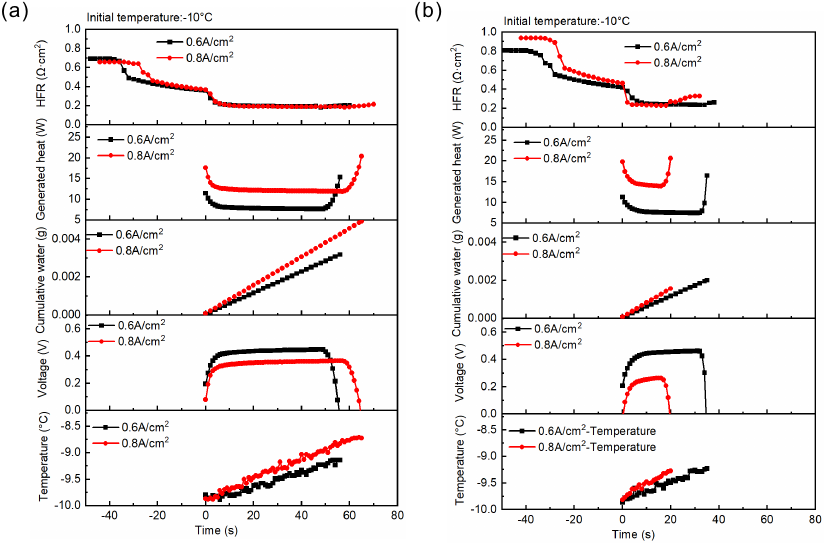}
  \caption{Time variation of parameters of the PEMFC under the constant current mode when the initial startup temperature is -10 $^\circ$C: (a) MFFF fuel cell; (b) SFF fuel cell. $t = 0$ in the figure corresponds to the instant when the electronic load is applied.}\label{fig:07}
\end{figure}
When the initial startup temperature is -10 $^\circ$C, both of the two fuel cells fail to start. For failure cases, we can use the operation time and final temperature rise to assess the startup performance. Here, the operation time denotes the time since the application of the load until the voltage drops to 0, and the final temperature rise refers to the increment of the temperature until the voltage drops to 0. According to Figure \ref{fig:07}, under the constant current modes of 0.6 and 0.8 A/cm$^2$, the output voltage of the fuel cells rises slowly for a period of time, then suddenly drops sharply until the voltage reversal occurs. The process exhibits three distinct stages according to the variation in the HFR. After the cold start begins, the HFR declines rapidly, while with the progress of the reaction, the HFR declines slowly for a period of time. Then the HFR suddenly rises, indicating that ice appears inside the PEMFC. The ice formation leads to the coverage of CL reaction sites and also hinders the gas transfer. The instant when the output voltage suddenly drops is basically the same as the instant when the HFR suddenly rises. According to Figure \ref{fig:07}a, the final temperature rise and operation time of the MFFF fuel cell at 0.8 A/cm$^2$ are larger than those at 0.6 A/cm$^2$. According to Figure \ref{fig:07}b, the final temperature of the SFF fuel cell at 0.6 and 0.8 A/cm$^2$ is basically the same, but the operation time at 0.6 A/cm$^2$ is longer than at 0.8A/cm$^2$. This difference indicates that the SFF fuel cell exhibits an accelerated water production rate when subjected to higher current densities, and leads to faster ice formation and quicker failure of the startup than under lower current densities. Comparing the two fuel cells, their water production rates are almost the same. However, the porous structure of the MFFF has more channels to remove the water, hence resulting in a prolonged operation time compared with the SFF fuel cell.

The cold start time of the MFFF and SFF fuel cells under constant current mode are summarized in Table~\ref{tab:01}. By summarizing the results in the constant current mode, we can find that the rates of heat production and water production are different under different current densities. The icing rate is correlated with the rate at which water is produced. An increase in current density is directly associated with an increase in voltage loss and a decrease in output voltage. Additionally, higher heat release occurs, resulting in a more rapid increase in temperature at elevated current densities. The cold start of the two PEMFCs exhibits differences at the initial startup temperature of -10 $^\circ$C. The MFFF fuel cell exhibits enhanced cold start characteristics at higher current density in comparison to lower current density. The operation time of the SFF fuel cell is seen to be longer when operating at a lower current density compared to a higher current density. The water production of the SFF fuel cell is smaller at lower current density than at higher current density. A higher current density will result in the acceleration of the water and the occurrence of water freezing at an earlier stage. When the initial startup temperature is -5 $^\circ$C and -7 $^\circ$C, it is evident that the MFFF fuel cell exhibits a greater operating voltage in comparison to the SFF fuel cell. Hence, the heat production rate of the SFF fuel cell surpasses that of the MFFF fuel cell. Meanwhile, the water production rate of the PEMFCs is the same. Therefore, the SFF fuel cell reaches 0 $^\circ$C in a shorter time than the MFFF fuel cell. When the initial startup temperature is -10 $^\circ$C, the operation time of the MFFF fuel cell is extended compared with that of the SFF fuel cell. This can be primarily attributed to the superior drainage effect of the MFFF fuel cell compared with the SFF fuel cell.

\begin{table}[]
\centering
\caption{Cold start time of the MFFF and SFF fuel cells under constant current mode.}
\label{tab:01}
\begin{tabular}{ccccc}
\hline
Temperature        & MFFF fuel cell        & MFFF fuel cell       & SFF fuel cell       & SFF fuel cell                \\
($^\circ$C)        & 0.6 (A/cm$^2$)        & 0.8 (A/cm$^2$)       & 0.6 (A/cm$^2$)      & 0.8 (A/cm$^2$)               \\ \hline
-5                 & 616 s                 & 369 s                & 609 s               & 360 s                        \\
-7                 & 1225 s                & 703 s                & 1140 s              & 634 s                        \\
-10                & Failed at 56 s        & Failed at 65 s       & Failed at 35 s      & Failed at 20 s               \\ \hline
\end{tabular}
\end{table}

\subsubsection{Constant voltage mode}\label{sec:3.2.2}
\begin{figure}
  \centering
  \includegraphics[width=\columnwidth]{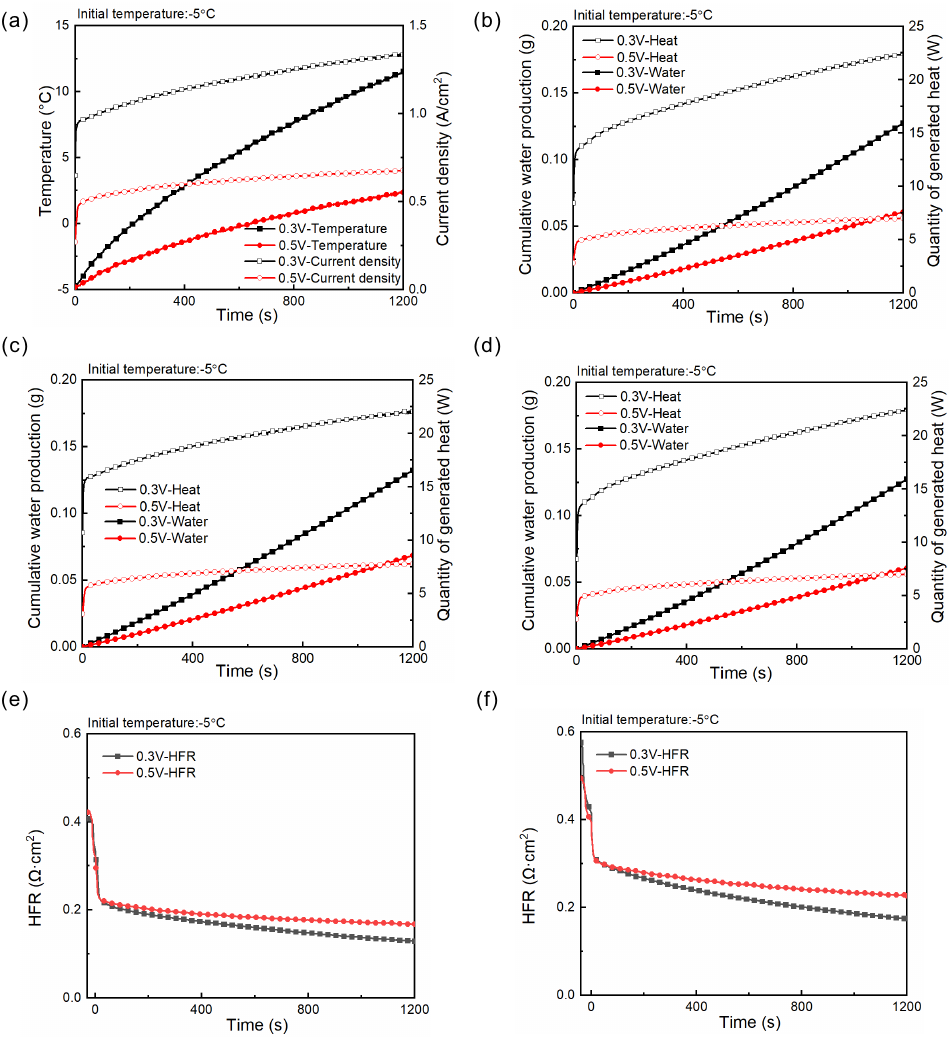}
  \caption{Time variation of parameters of the PEMFC under the constant voltage mode when the initial startup temperature is -5 $^\circ$C: (a, c, e) MFFF fuel cell; (b, d, f) SFF fuel cell. $t = 0$ in the figure corresponds to the instant when the electronic load is applied.}\label{fig:08}
\end{figure}

We tested the cold start of PEMFCs under constant voltage mode at 0.3 and 0.5 V, which are often used for the cold start of fuel cells in the constant voltage mode \cite{jiao11, WOS:000594124400002, WOS:000355884300015, WOS:000370306300096, WOS:000336696100021}. A voltage of 0.5 V is typically used as a standard load for fuel cells during normal operation, while a voltage of 0.3 V generates more heat in the fuel cell, which is beneficial for cold start. The experimental results show that, when the initial startup temperature is -5 $^\circ$C, the two fuel cells both start successfully at the constant voltage mode. The gradual increase in both temperature and current density is evident from Figures \ref{fig:08}a and \ref{fig:08}b. This may be attributed to the increasing heat generated. Meanwhile, as the temperature of the PEMFCs rises, the conductivity of the membrane, the rate of electrochemical reaction, and the current density increase, albeit at a gradual pace. The increase rate of the temperature and the current density gradually decreases, mainly because the ohmic resistance gradually decreases and the heat production rate of the PEMFCs decreases. The HFR of both fuel cells gradually decreases as observed in Figures \ref{fig:08}e and \ref{fig:08}f, suggesting that the heat produced by the fuel cells is sufficient to melt the generated ice at sub-freezing temperature so that the reaction can proceed smoothly. The current of the MFFF fuel cell is higher than that of the SFF fuel cell when operating at a constant voltage. According to Eq.~(\ref{eq:01}), the heat production rate of the MFFF fuel cell surpasses that of the SFF fuel cell. Hence, the rate of temperature rise of the MFFF fuel cell is comparatively higher.

\begin{figure}
  \centering
  \includegraphics[width=\columnwidth]{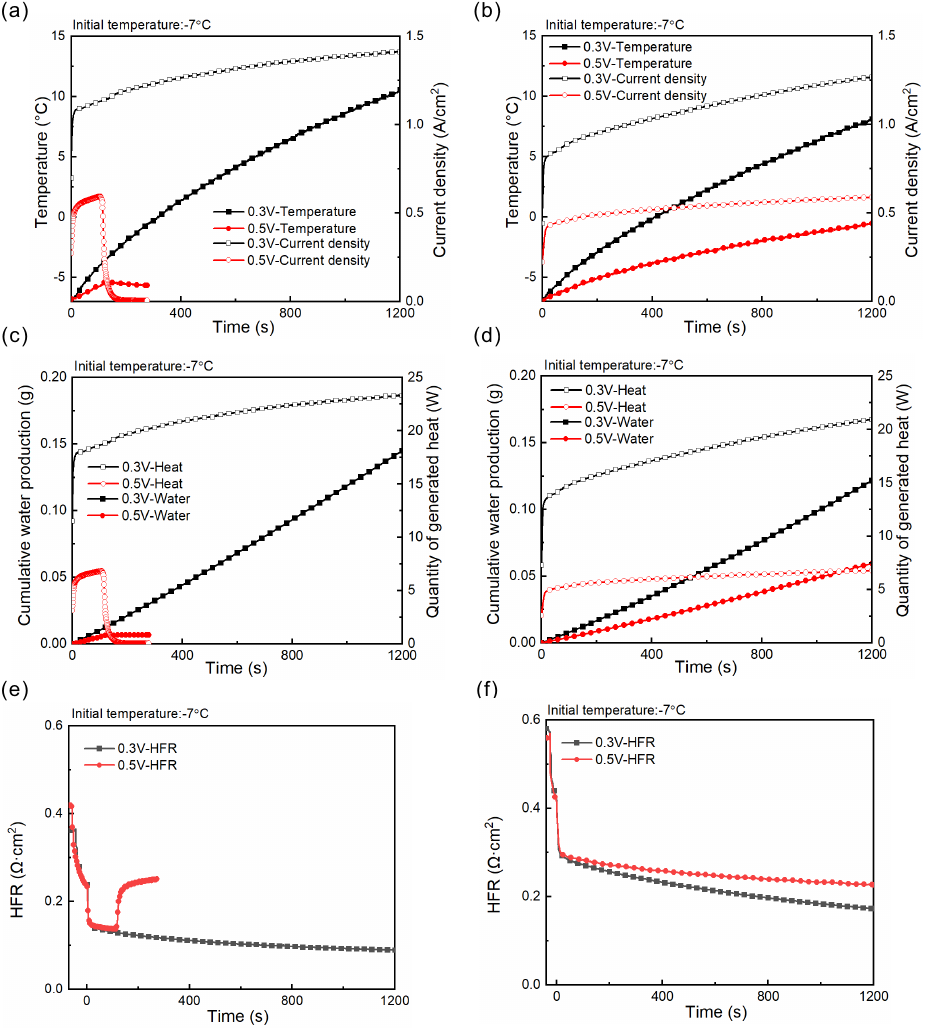}
  \caption{Time variation of parameters of the PEMFC in the constant voltage mode when the initial startup temperature is -7 $^\circ$C: (a, c, e) MFFF fuel cell; (b, d, f) SFF fuel cell. $t = 0$ in the figure corresponds to the instant when the electronic load is applied.}\label{fig:09}
\end{figure}

At the initial startup temperature of -7 $^\circ$C, the MFFF and SFF fuel cells both start successfully at 0.3 V, as depicted in Figure \ref{fig:09}. The MFFF fuel cell fails to start at 0.5 V, and the current exhibited a gradual decline at 105 s, ultimately reaching 0.02 A/cm$^2$ at 159 s. Meanwhile, the HFR of the MFFF fuel cell suddenly rises. The observed change in the HFR indicates that after the water exceeds the capacity, the product water starts to freeze gradually in the CL pores. Eventually, the reaction rate gradually decreases until ice particles completely cover the CL. Then, the electrochemical reactions cease, and the cold start fails.

\begin{figure}
  \centering
  \includegraphics[width=\columnwidth]{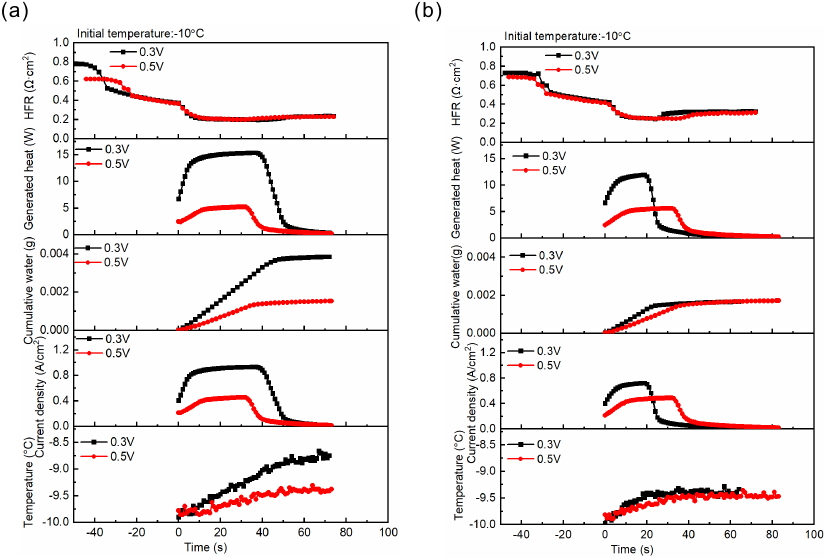}
  \caption{Time variation of parameters of the PEMFC in the constant voltage mode when the initial startup temperature is -10 $^\circ$C: (a) MFFF fuel cell; (b) SFF fuel cell. $t = 0$ in the figure corresponds to the instant when the electronic load is applied.}\label{fig:10}
\end{figure}

When the initial startup temperature is -10 $^\circ$C, both fuel cells fail to start at 0.3 and 0.5 V, as depicted in Figure \ref{fig:10}. When the MFFF fuel cell is subjected to an initial startup voltage of 0.3 V, it exhibits enhanced performance in terms of higher current densities, resulting in more water production and a lower minimum value of HFR than that at 0.5 V. The operation time of the MFFF fuel cell is basically the same at different startup voltages, but the final temperature rise of the MFFF fuel cell is greater at 0.3 V than at 0.5 V. This indicates that decreasing the voltage is beneficial to improve the performance of the MFFF fuel cell. The SFF fuel cell operates for a longer period at 0.5 V and eventually produces more water, resulting in a lower value of HFR than at 0.3 V. The MFFF fuel cell operates at a greater current density and produces more water than the SFF fuel cell, but the MFFF fuel cell operates for a longer period, which is mainly due to the better drainage effect of the MFFF fuel cell in comparison to the SFF fuel cell at 0.3 V. At 0.5 V, the MFFF fuel cell has a shorter operation time than the SFF fuel cell due to the higher thermal conductivity of the metal foam relative to graphite, and the rate of heat loss from the MFFF fuel cell is faster than that of the SFF fuel cell, leading to an accelerated icing rate in comparison to the SFF fuel cell.

The cold start time of the MFFF and SFF fuel cells under constant voltage mode are summarized in Table~\ref{tab:02}.
By summarizing the results in the constant voltage mode, we can see that the current mainly depends on the applied load voltage. A lower load voltage results in a higher current. After applying the constant voltage load, the output current shows an increase and then a decrease, with a large slope of increase and decrease. This is mainly because the pulse current corresponding to the voltage load is large when the load is applied, so the water production rate in the initial period is fast, and the hydration of the membrane is also fast. In terms of the HFR variation, the HFR decreases rapidly in the initial stage. If the cold start is successful, the output current then shows a trend of slowly increasing until the temperature reaches 0 $^\circ$C, and the HFR decreases gradually. In the event of a failed cold start, the current suddenly decreases after gradually increasing, and the HFR suddenly increases after slowly decreasing. When the initial startup voltage is 0.3 V, the current of the MFFF fuel cell is higher than that of the SFF fuel cell when operating at a constant voltage. The heat production rate of the MFFF fuel cell surpasses that of the SFF fuel cell. Hence, the rate of temperature rise of the MFFF fuel cell is comparatively higher.

\begin{table}[]
\centering
\caption{Cold start time of the MFFF and SFF fuel cells under constant voltage mode.}
\label{tab:02}
\begin{tabular}{ccccc}
\hline
Temperature        & MFFF fuel cell        & MFFF fuel cell       & SFF fuel cell       & SFF fuel cell                \\
($^\circ$C)        & 0.3 V                 & 0.5 V                & 0.3 V               & 0.5 V                        \\ \hline
-5                 & 215 s                 & 647 s                & 275 s               & 740 s                        \\
-7                 & 320 s                 & Failed at 159 s      & 411 s               & 1451 s                       \\
-10                & Failed at 72 s        & Failed at 73 s       & Failed at 64 s      & Failed at 83 s               \\ \hline
\end{tabular}
\end{table}

\subsubsection{Ramping current mode}\label{sec:3.2.3}
\begin{figure}
  \centering
  \includegraphics[width=\columnwidth]{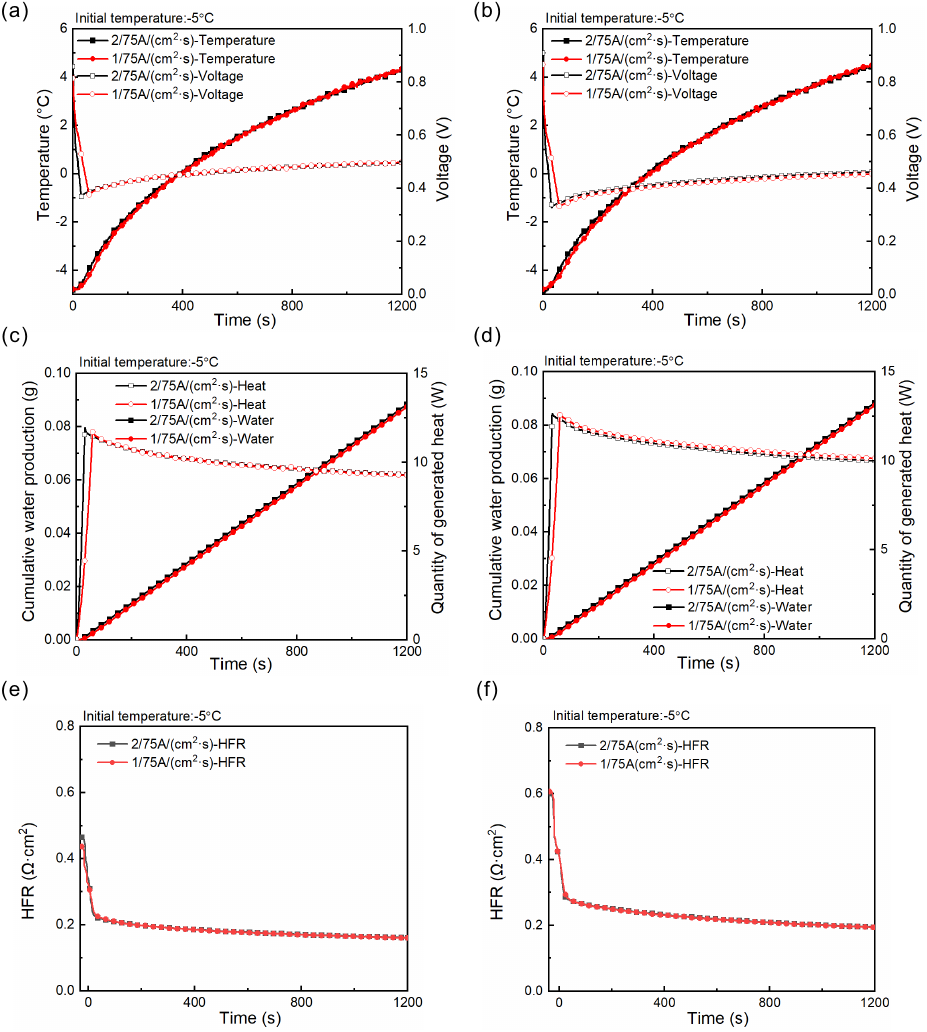}
  \caption{Time variation of parameters of the PEMFC in the ramping current mode when the initial startup temperature is -5 $^\circ$C: (a, c, e) MFFF fuel cell; (b, d, f) SFF fuel cell. $t = 0$ in the figure corresponds to the instant when the electronic load is applied.}\label{fig:11}
\end{figure}

We tested the ramping current mode of cold start, where the current density is raised linearly from 0 to a certain value at a constant rate, and then the current density is held constant. We tested two settings of the ramping current mode, where the current density is raised linearly from 0 to 0.8 A/cm$^2$ at the rates of 1/75 and 2/75 A/(cm$^2$$\cdot$s), respectively, and then the current density is maintained at 0.8 A/cm$^2$. The rates of temperature rise of the PEMFCs in the ramping current mode are all slightly lower than that at 0.8 A/cm$^2$ constant current mode at the initial startup temperature of -5 $^\circ$C. As depicted in Figures \ref{fig:11}a and \ref{fig:11}b, the decrease in voltage can be attributed to the increase in the applied current density during the initial stage. Subsequently, the voltage begins to gradually increase once the current density reaches 0.8 A/cm$^2$, aligning with the observed trend of the voltage under the constant current mode during this stage. The rates of heat and water production are basically the same under the two ramping current modes, and the decrease rate of the HFR is approximately the same.

\begin{figure}
  \centering
  \includegraphics[width=\columnwidth]{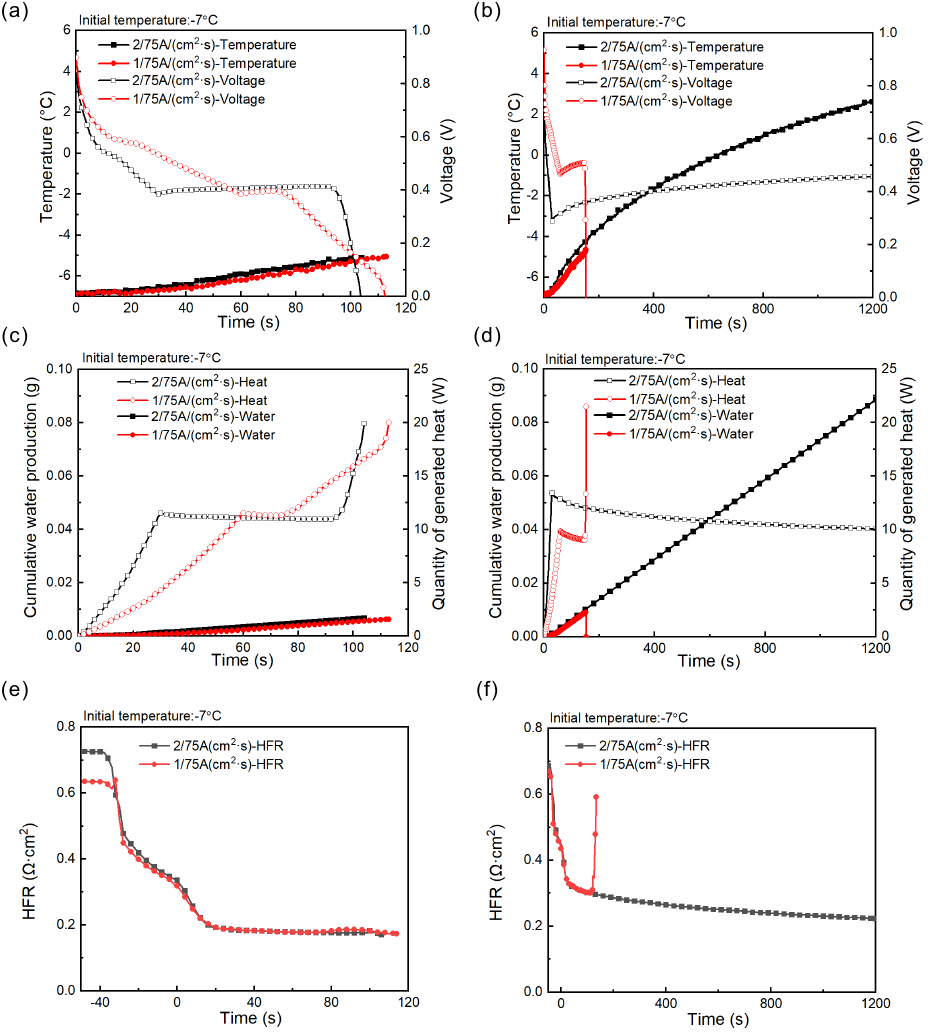}
  \caption{Time variation of parameters of the PEMFC in the ramping current mode when the initial startup temperature is -7 $^\circ$C: (a, c, e) MFFF fuel cell; (b, d, f) SFF fuel cell. $t = 0$ in the figure corresponds to the instant when the electronic load is applied.}\label{fig:12}
\end{figure}

At the initial startup temperature of -7 $^\circ$C, the startup of the MFFF fuel cell failed under different ramping current modes, as depicted in Figure \ref{fig:12}a. The voltage of the MFFF fuel cell reverses at 104 s at 2/75 A/(cm$^2\cdot$s), while the voltage reverses at 113 s at 1/75 A/(cm$^2\cdot$s). The SFF fuel cell started successfully at 2/75 A/(cm$^2\cdot$s), and took a total of 636 s to achieve  0$^\circ$C. However, the voltage of the SFF fuel cell started to plummet and the electrochemical reaction stopped at 154 s at 1/75 A/(cm$^2\cdot$s).

\begin{figure}
  \centering
  \includegraphics[width=\columnwidth]{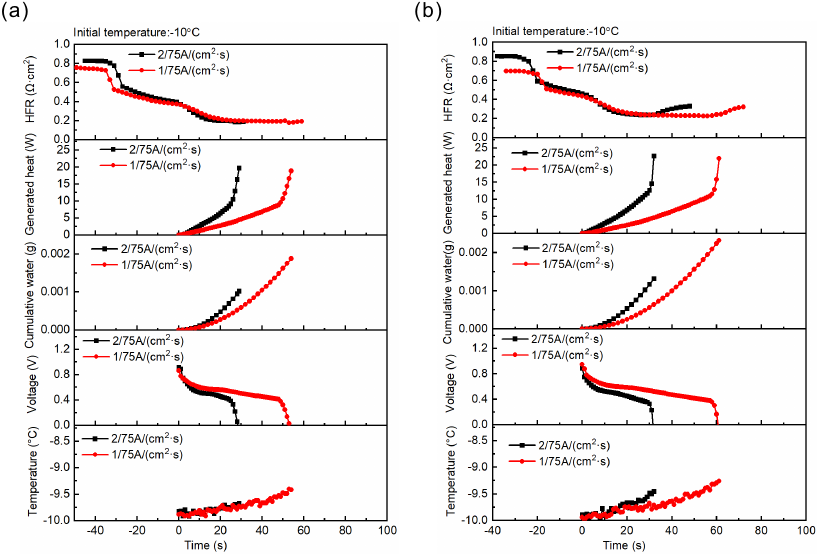}
  \caption{Time variation of parameters of the PEMFC in the ramping current mode when the initial startup temperature is -10 $^\circ$C: (a) MFFF fuel cell; (b) SFF fuel cell. $t = 0$ in the figure corresponds to the instant when the electronic load is applied.}\label{fig:13}
\end{figure}

The performance of the MFFF and SFF fuel cells at the initial startup temperature of -10 $^\circ$C is shown in Figure \ref{fig:13}. As depicted in Figure \ref{fig:07}, the MFFF fuel cell has an operation time of 65 s and a final temperature of -8.7 $^\circ$C at 0.8 A/cm$^2$, and the SFF fuel cell has an operation time of 20 s and a final temperature of -9.3 $^\circ$C at 0.8 A/cm$^2$. Compared with the operation time and final temperature rise of fuel cells under the ramping current mode in Figure \ref{fig:13}, we can see that the ramping current mode has no significant improvement on the operation time and final temperature rise of the MFFF fuel cell at -10 $^\circ$C. The final temperature rise of the SFF fuel cell at 1/75 A/(cm$^2\cdot$s) is basically the same as that of constant current mode at 0.8 A/cm$^2$, but the fuel cell lasts longer at 1/75 A/(cm$^2\cdot$s) than that at 0.8 A/cm$^2$. At the current increase rate of 2/75 A/(cm$^2\cdot$s), the rates of water and heat production of the MFFF and SFF fuel cells are larger than that at 1/75 A/(cm$^2\cdot$s), but the operation time is shorter than at 1/75 A/(cm$^2\cdot$s).

The cold start time of the MFFF and SFF fuel cells under ramping current mode are summarized in Table~\ref{tab:03}.
For the ramping current mode, at -5 $^\circ$C and -7 $^\circ$C, the performance of the MFFF and SFF fuel cells at two different current increase rates did not exhibit superior to that at constant current modes. This is mainly because the higher current under the constant current mode contributes to the success of the startup at -5 $^\circ$C and -7 $^\circ$C, whereas the ramping current mode operates at low densities for a certain duration. At -10 $^\circ$C, under the ramping current mode, the operation time and final temperature rise of the MFFF fuel cell are lower than those at 0.8 A/cm$^2$. However, the SFF fuel cell at 1/75 A/(cm$^2\cdot$s) has a longer operation time than that at constant current modes, indicating that selecting the appropriate initial current density and slope can improve the cold start. When the initial startup temperature is -10 $^\circ$C, under the ramping current mode, the operation time and final temperature rise of the MFFF fuel cell are lower than those at 0.8 A/cm$^2$. However, the SFF fuel cell at 1/75 A/(cm$^2\cdot$s) has a longer operation time than that at constant current modes, indicating that selecting the appropriate initial current density and slope can improve the performance of PEMFCs.

\begin{table}[]
\centering
\caption{Cold start time of the MFFF and SFF fuel cells under ramping current mode.}
\label{tab:03}
\begin{tabular}{ccccc}
\hline
              & MFFF fuel cell              & MFFF fuel cell                      & SFF fuel cell                 & SFF fuel cell                             \\
Temperature   & 0.8 (A/cm$^2$)              & 0.8 (A/cm$^2$)                      & 0.8 (A/cm$^2$)                & 0.8 (A/cm$^2$)                            \\
($^\circ$C)   & -1/75 A/(cm$^2\cdot$s)      & -1/75 A/(cm$^2\cdot$s)              &-1/75 A/(cm$^2\cdot$s)         & -1/75 A/(cm$^2\cdot$s)                    \\ \hline
-5            & 393 s                       & 400 s                               & 384 s                         & 394 s                                     \\
-7            & Failed at 104 s             & Failed at 113 s                     & 636 s                         & Failed at 154 s                           \\
-10           & Failed at 29 s              & Failed at 54 s                      & Failed at 32 s                & Failed at 61 s                            \\ \hline
\end{tabular}
\end{table}

\subsubsection{Variable current mode}\label{sec:3.2.4}
Generally, the cold start procedure of fuel cells includes three distinct stages \cite{wan14}: hydration stage, undersaturated stage, and melting stage \cite{gwak15}. Initially, the water generated by the Oxygen Reduction Reaction (ORR) primarily serves to hydrate the membrane, and the HFR experiences a significant decline at this stage. As the reaction continues, the generated water vapor may freeze at the cathode CL once it reaches freezing temperature. Meanwhile, the generated heat will heat the PEMFCs. In the event that the cathode CL becomes entirely covered by ice prior to the temperature reaching 0 $^\circ$C, the reaction stops and the cold start fails. If the fuel cell can continue operating to above 0 $^\circ$C, the ice starts to melt and the cell continues to heat up to operating temperature \cite{wang10}. In the experiment of this study, it can be seen that the hydration stage is from the application of load to 10 s, and then the unsaturated stage is before the temperature reaches 0 $^\circ$C, and the melting stage is after the successful cold start. During different stages, the water production, heat production, and icing are different. Therefore, different load currents can be applied at different stages to achieve an optimal balance between heat production and water production. This approach is expected to effectively mitigate the occurrence of internal icing. For this purpose, six different settings of variable current startup mode are designed, as listed in Table \ref{tab:04}. The current remains constant for a certain amount of time, then increases (or decreases) at a constant rate, and then remains constant. For example, Case 1 is where the current density is held at 0.6 A/cm$^2$ for 10 s, then increases to 0.8 A/cm$^2$ at a constant slope of 1/50 A/(cm$^2\cdot$s) and then remains at 0.8 A/cm$^2$. Case 1 is to increase the current in the hydration stage, Case 2 is to increase the current in the unsaturated stage, Case 3 is to reduce the current in the hydration stage, Case 4 is to reduce the current in the unsaturated stage, and Case 5 and 6 are linear slope current modes, which differ in the rate of decrease.

\begin{table}[]
\centering
\caption{Parameters for variable current mode.}
\label{tab:04}
\begin{tabular}{ccccc}
\hline
       & Initial current density & Operation time & Slope              & Final current density        \\
       & (A/cm$^2$)              & (s)            & (A/(cm$^2\cdot$s)) & (A/cm$^2$)                   \\ \hline
Case 1 & 0.6                     & 10             & 1/50               & 0.8                          \\
Case 2 & 0.6                     & 30             & 1/50               & 0.8                          \\
Case 3 & 0.8                     & 10             & -1/50              & 0.6                          \\
Case 4 & 0.8                     & 30             & -1/50              & 0.6                          \\
Case 5 & 0.8                     & -              & -1/150             & 0.6                          \\
Case 6 & 0.8                     & -              & -1/300             & 0.6                          \\ \hline
\end{tabular}
\end{table}

The variable current mode was tested at an initial startup temperature of -10 $^\circ$C. Generally, the temperature increase of fuel cells is expected to be greater at a higher current. However, at an initial startup temperature of -10 $^\circ$C, as the current density increases, the rate of water production also increases, which can lead to freezing at low temperatures, causing the fuel cell to function for only a short period of time and resulting in a failure of cold start. Thus, we tested the variable current mode at a low initial startup temperature of -10 $^\circ$C. Then, in the variable current mode, the fuel cell can operate initially in at a relatively lower current for a specific duration, which prolong the continuous operation of the fuel cell, but does not produce much water for icing. In contrast, if the initial startup temperature is not low enough, such as at -5 $^\circ$C and -7 $^\circ$C, the operation at relatively lower current is not necessary, and a high current density (e.g., 0.8 A/cm$^2$) can be applied directly.

\begin{figure}
  \centering
  \includegraphics[width=\columnwidth]{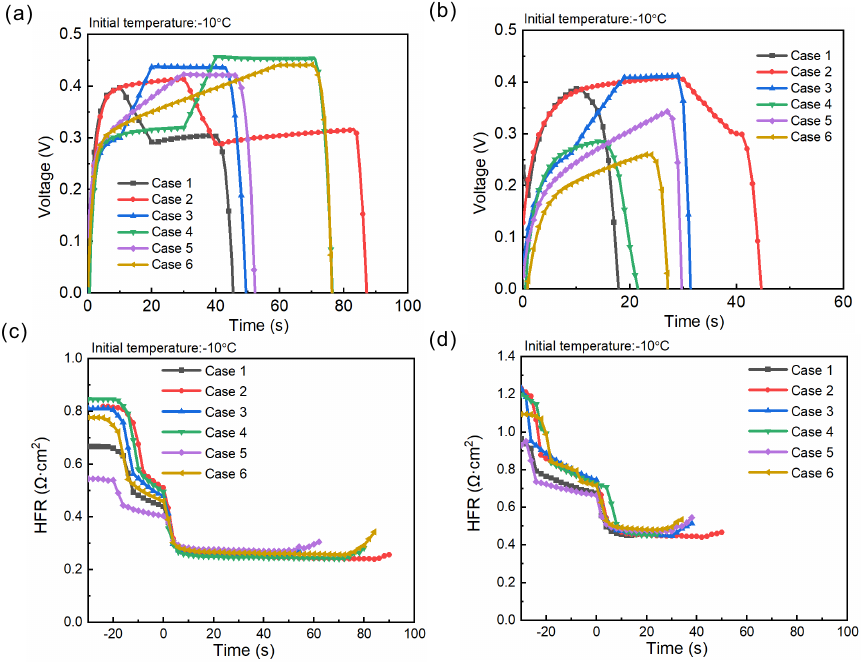}
  \caption{Changes of voltage and HFR in the variable current mode at the initial startup temperature of -10 $^\circ$C: (a, c) MFFF fuel cell; (b, d) SFF fuel cell. $t = 0$ in the figure corresponds to the instant when the electronic load is applied.}\label{fig:14}
\end{figure}

The variation of the HFR and voltage of the MFFF fuel cell and SFF fuel cell in the variable current startup mode are depicted in Figure \ref{fig:14}. Under the variable current mode, the voltage of the fuel cell increases in a relatively short period of time, and the extent of this increase depends on the initial current applied. As the reaction proceeds, the quantity of water progressively rises, while the heat generation gradually fails to completely melt the solid ice generated. When the reaction fails to proceed, the output voltage drops to zero in a very short time, representing a cold start failure.

\begin{figure}
  \centering
  \includegraphics[width=\columnwidth]{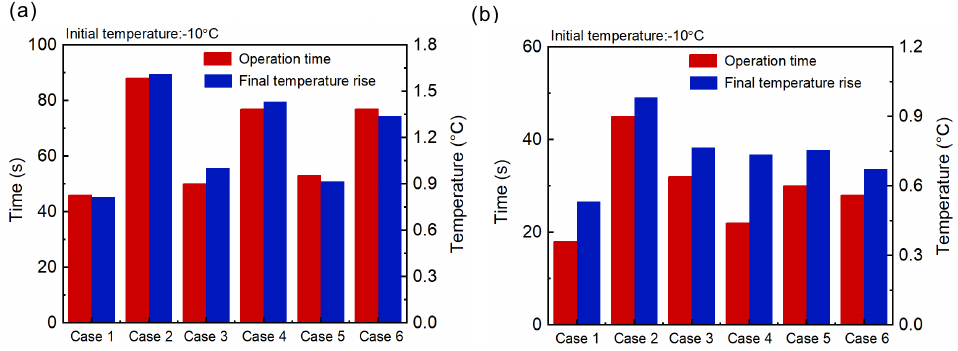}
  \caption{Operation time and final temperature rise in the variable current mode at the initial startup temperature of -10 $^\circ$C: (a) MFFF fuel cell; (b) SFF fuel cell.}\label{fig:15}
\end{figure}

The operation time and final temperature rise of the MFFF fuel cell and SFF fuel cell under variable current modes are depicted in Figure \ref{fig:15}. Compared with the constant current mode, the performance of the MFFF fuel cell and SFF fuel cell under Case 1 decreases, the operation time is shortened, and the final temperature rise decreases, indicating that increasing the current in the hydration stage will accelerate the formation and accumulation of ice. The MFFF fuel cell and SFF fuel cell in Case 2 has superior performance compared with other startup modes. The operation time of the MFFF fuel cell is 88 s, and the final temperature rise of the MFFF fuel cell is 1.6 $^\circ$C, while the operation time of the SFF fuel cell is 46 s, and the final temperature rise of the MFFF fuel cell is 0.98 $^\circ$C. This indicates that increasing the current in the unsaturated stage can increase heat production and expedite the temperature rise, while preventing additional accumulation of ice.

\begin{figure}
  \centering
  \includegraphics[width=\columnwidth]{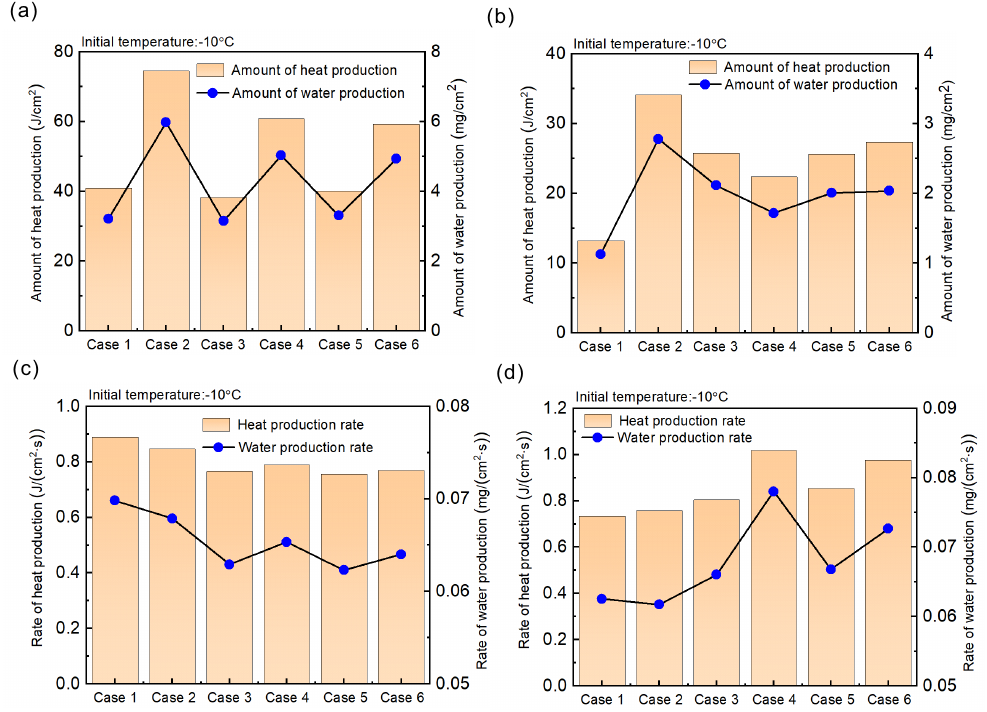}
  \caption{Water and heat production under different cases in the variable current mode: (a, c) MFFF fuel cell; (b, d) SFF fuel cell.}\label{fig:16}
\end{figure}
Heat production and water production are two key factors affecting the cold start. The heat production and water production of the MFFF fuel cell under variable current modes are illustrated in Figure \ref{fig:16}. From Figures \ref{fig:16}a and \ref{fig:16}b, it is evident that among the different variable current modes, Case 2 has the largest water production and the most significant heat production of the MFFF fuel cell and the SFF fuel cell, whereas Case 3 has the least total heat production and water production of the MFFF fuel cell, and Case 1 has the least total heat production and water production of the SFF fuel cell. However, the fuel cell under the six variable current modes has different operation times, so it is more appropriate to use the rates of heat production and water production to compare the different variable current modes. The rates of heat production and water production in different variable current modes are shown in Figure \ref{fig:16}b and \ref{fig:16}d. Case 1 has the shortest operation time of the MFFF fuel cell, and its rates of heat production and water production are the largest. The rates of heat production and water production of the SFF fuel cell are the largest in Case 4. Case 2 has a higher rate of heat production and a comparatively lower rate of water production, which is the best among the six variable current modes.

\section{Conclusions}\label{sec:4}
In this paper, the cold start of MFFF and SFF fuel cells is studied experimentally, and the effects of startup modes are investigated. We focus on the output voltage, current density, maximum operation time, water production, heat production, and HFR variation. The following conclusions are drawn:
\begin{itemize}
  \item Under the constant voltage mode, the cold start performance exhibits an improvement as the initial voltage decreases.
  \item Under the constant current mode, the MFFF fuel cell has a faster rate of temperature increase in high currents than in low currents.
  \item For the SFF fuel cell, enhancing the current at temperatures of -5 $^\circ$C and -7 $^\circ$C leads to an improvement in the performance. After decreasing the initial startup temperature, the temperature increase of the SFF fuel cell can be enhanced with a high current density, and a low current density can extend the maximum operation time of the SFF fuel cell. When the initial startup temperature is -10 $^\circ$C, the implementation of the ramp current mode can extend the operation time of the SFF fuel cell in the condition of the same final temperature.
  \item The variable current mode is designed based on the heat production and water production at different stages. The findings indicate that increasing the current at the unsaturated stage leads to an elevated heat production rate and a reduced water production rate, which can improve the performance of the MFFF fuel cell. Under the variable current mode, the MFFF fuel cell experiences a minimum increase of 22 seconds in operation time and a 26\% increase in temperature rise.
\end{itemize}

\section*{Acknowledgements}
This work was financially supported by the National Natural Science Foundation of China (Grant Nos. 51920105010, and 51921004) and the Department of Science and Technology of Inner Mongolia Autonomous Region (Grant No. 2022JBGS0027).

\bibliography{ColdStartMode}

\end{document}